# Research Entity Extraction and Topic Detection from UKRI Grant Proposals

Xingran Ruan[1], Angelo Salatino[2], Rosa Filgueira[1], Kara Moraw[1], Alexandru Marcoci[3], Gemma Derrick[4], and Sarah Callaghan[5]

[1] *x.ruan@epcc.ed.ac.uk; r.filgueira@epcc.ed.ac.uk; k.moraw@epcc.ed.ac.uk*
0000-0001-9462-8816; 0000-0002-5715-3046; 0009-0005-2202-7006
EPCC, University of Edinburgh, UK

[2] *angelo.salatino@open.ac.uk*
0000-0002-4763-3943
Knowledge Media Institute, The Open University, UK

[3] *am3159@cam.ac.uk*
0000-0002-5780-0805
Institute for Technology and Humanity, University of Cambridge, UK

[4] *gemma.derrick@bristol.ac.uk*
0000-0001-5386-8653
Centre for Higher Education Transformations, School of Education, University of Bristol, UK

[5] *sarah.callaghan@admin.ox.ac.uk*
0000-0002-0517-1031
University of Oxford, UK

This paper presents preliminary findings from a UKRI-funded Metascience project comparing three LLM-based approaches, GPT-4o, Mistral, and a bespoke algorithm, DSIT-Taxonomies, for extracting and classifying research entities from funding proposals. Our project "Tracking Stars and Unicorns" aims to identify early signals of emerging research areas to inform public investment. Our methodology employed a three-stage pipeline, leveraging Mistral for primary entity extraction and mapping against the OpenAlex Topics taxonomy. We evaluated our approach across 42 proposals' abstracts from different areas and observed that Mistral and GPT-4o produce comparable, high-quality entity sets with significant semantic overlap, outperforming the fragmented DSIT-Taxonomies approach. Crucially, the Mistral-based approach achieved superior topic classification accuracy (90.5%) compared to the full DSIT-Taxonomies pipeline (71.4%). We conclude that Mistral offers a high-performance, operationally efficient, and secure solution for large-scale analysis of sensitive grant data.

## 1. Introduction

In the global knowledge economy, publicly competitively funded research plays a central role in driving innovation, national productivity and economic growth (Sussex et al, 2016). As such, governments are increasingly seeking to optimise returns on public research investment by identifying and prioritising areas of emerging strength. In the UK, this has translated into a growing policy emphasis of research specialisation (DfE, 2025), focusing resources on a smaller number of domains in line with national priorities, or alternatively, promoting the production of 'excellence' (Moore et al, 2017). It is argued that a policy of specialisation will enhance productivity, foster more innovative research ecosystems and strengthen international competitiveness (DfE, 2025).

Realising the benefit of research specialisation, however, depends on the ability to identify, and fund, promising areas at an early, formative stage or when funding is likely to accelerate the process of discovery. The generation of research topics, particularly within *ex-ante* funding proposals, i.e., research yet to be funded and yet to be done, offers a valuable but overlooked perspective for understanding how future areas of excellence emerge. Early-stage signals, phrases or concepts used in *ex-ante* proposal texts have the potential to indicate how 'unicorn' domains with future excellence potential and are likely to benefit from early and timely public investment to catalyse the realisation of significant downstream societal or economic impact/benefit (Yin et al, 2022). However, despite this potential, systematically extracting these signals at scale presents significant technical challenges.

Recent advances in large language models (LLMs) offer new tools capable of addressing these challenges and have already demonstrated considerable promise in tasks of entity extraction (Gartlehner et al, 2024), topic modelling (Aggarwal et al, 2025) and semantic analysis across large corpora of scientific text (Mansour et al, 2025). Despite this, the question remains in whether the reliability, consistency and comparability of different LLM approaches when applied to complex, domain-specific texts such as *ex-ante* research proposals.

This paper presents preliminary findings from “Tracking Stars and Unicorns”, a UKRI-funded Metascience project comparing how three LLM-based approaches, GPT-4o, Mistral and a bespoke algorithm developed by the Innovation Growth Lab (IGL) in collaboration with the UK Department of Science, Innovation and Technology (DSIT), in extracting and classifying research entities from grant proposals, called DSIT-Taxonomies (Innovation Growth Lab, 2025). By examining the degree of convergence/divergence across these three models, this research aims to assess their suitability for application in a larger corpus of ex-ante research proposals, and thus research that will contribute insights into the identification of ‘unicorns’ within research proposals. This research also contributes to broader debates on the role of automated methods in research policy and evaluation.

While the overarching project is based upon unprecedented access to a corpus of both funded and unfunded UKRI applications (approximately 350K), the evaluation in this paper is performed on a randomised set of 42 funded proposals’ titles and abstracts from seven UK statutory research councils and a specialist national centre. We intentionally focused our evaluation only on funded proposals because the outcome of a proposal should not be a factor in the quality of entities and topics extracted. Furthermore, funded proposal abstracts are openly available and can be shared with the community, unlike the abstracts of unfunded proposals, which are protected by copyright.

## 2. Related Work

### *2.1. Automated Extraction and Classification of Research Topics*

Early approaches for extracting research topics primarily relied on keyword co-occurrence and bibliometric mapping techniques using tools such as VOSviewer (Van Eck, Waltman 2010) and Bibliometrix (Aria et al., 2017). Subsequent ontology-driven systems, such as the CSO Classifier (Salatino et al., 2022), SciNoBo (Gialitsis et al., 2022), QuickUMLS (Soldaini & Goharian, 2016), and MetaMap (Aronson & Lang, 2010), introduced more structured annotation by mapping text to predefined taxonomies. However, these systems are often domain-specific and require substantial effort to apply across disciplines. Their reliance on fixed vocabularies also limits their ability to capture emerging terminology, particularly in *ex-ante* funding proposals where novel ideas precede established labels (Salatino et al., 2022).

In contrast, machine learning approaches, particularly transformer-based architectures like BERT (Devlin et al., 2019) and SciBERT (Beltagy et al., 2019) have demonstrated strong performance in scientific text analysis, through contextual embeddings. Building on these, recent LLMs offer superior semantic reasoning and broader cross-domain generalisation. Unlike ontology-driven systems, LLMs support entity extraction and zero-shot classification without relying on fixed vocabularies. Pretrained on vast scientific corpora, they are capable of interpreting newly emerging concepts, not yet formalised in existing taxonomies. These properties make LLM-based approaches for identifying early conceptual signals in emerging research areas.

*2.2. Knowledge Organisation Systems for Mapping Scientific Fields*

Knowledge Organisation Systems (KOSs) vary substantially in scope, depth, and intended use, ranging from domain-specific systems that cover a single field to broad taxonomies designed to span multiple disciplines. Selecting an appropriate KOS depends on specific needs for coverage, granularity, and interoperability (Salatino et al., 2025). Among general-purpose taxonomies, OpenAlex Topics is widely recognised for its comprehensiveness and adoption (Priem et al., 2022). Its four-level hierarchy, spanning domains, fields, subfields, and topics, enables multi-level analysis across disciplinary boundaries while capturing emerging research areas (Culbert et al., 2025). Furthermore, its role as a standardised reference framework in bibliometrics ensures interoperability and facilitates comparison across studies (Salatino et al., 2025). Consequently, the OpenAlex Topics was adopted to ensure seamless integration with downstream analytical workflows.

## 3. Materials and Methods

The workflow follows a modular three-stage pipeline to convert unstructured text within grant applications into structured data (Figure 1). The process initiates with the Prompt Generation module, ingesting raw Funding Proposals and serialises into a structured format using a predefined Prompt Template. These formatted prompts are then dispatched to an External LLM API to perform the Entity Extraction task. To ensure thematic alignment, the raw entities are mapped against the OpenAlex Topics. The final output consists of validated relevant entities and topics.

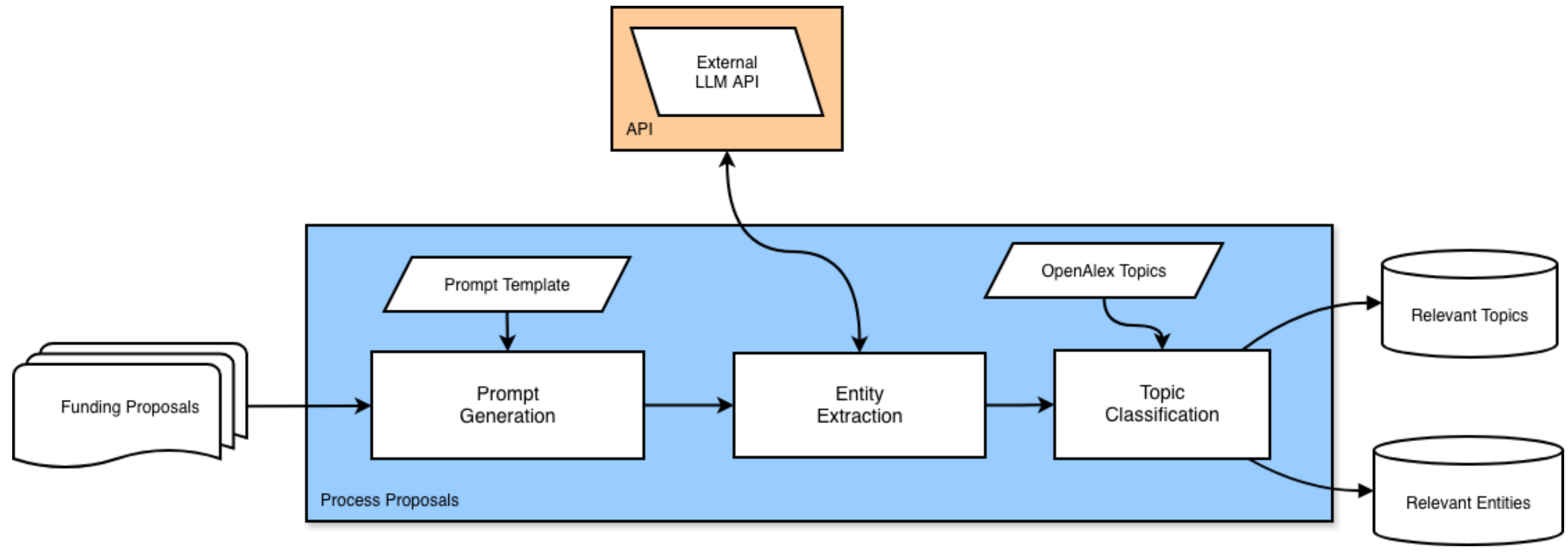

**Figure 1: Workflow of the Research Entity and Topic Extraction Pipeline.**

*3.1. Data Source*

The evaluation dataset was sourced from the UKRI's Gateway to Research (GtR) portal, a repository of publicly funded research project information. The dataset includes project titles, technical abstracts, and metadata such as funding value and lead organisation. An ad-hoc crawler facilitated systematic data collection from individual research councils, resulting in a custom-built database. The proposals (*n=42*) span the full breadth of the disciplinary landscape in the UK, spanning the seven statutory research councils and a specialist national centre. These include:

1. *Arts and Humanities Research Council* (AHRC, *n=5*), which supports research ranging from history to the creative industries,
2. *Biotechnology and Biological Sciences Research Council* (BBSRC, *n=5*), which invests in bioscience to deliver sustainable bio-based solutions,
3. *Medical Research Council* (MRC, *n=5*), dedicated to improving human health through medical science,

4. *Economic and Social Research Council* (ESRC, *n=6*), which acts as the primary funder for economic, social, and human data science,
5. *Natural Environment Research Council* (NERC, *n=5*), covering environmental sciences,
6. *Science and Technology Facilities Council* (STFC, *n=5*) focusing on fundamental physics, astronomy, and nuclear science,
7. *Engineering and Physical Sciences Research Council* (EPRSC, *n=6*) provides the data for the foundations of engineering, mathematics, and chemistry, and, finally
8. *National Centre for the Replacement, Refinement and Reduction of Animals in Research* (NC3Rs, *n=5*), a specialist scientific organisation primarily funded by the MRC and the BBSRC.

*3.2. Extracting Research Entities*

To map the conceptual landscape of the research proposals, this study employed an automated entity extraction pipeline leveraging the Mistral language model (Jiang et al, 2023). This process transitioned raw text into a structured knowledge representation by identifying and categorising core scientific components. In this context, a *research entity* is defined as a discrete semantic unit representing a functional component of a scientific inquiry, including thematic topics, disciplines, methodologies, and technical facets such as software and datasets. Furthermore, this taxonomy incorporates tangible artefacts and significant historical or environmental events. The extraction of these occurrences is crucial for a longitudinal analysis, as it provides the necessary temporal grounding to correlate shifting research trends with specific external catalysts. By identifying events such as the COVID-19 pandemic or the 2008 financial crisis, the model transforms static keywords into dynamic indicators, revealing how UK funding bodies pivot research priorities in response to global or domestic shocks.

The extraction was achieved through an iterative prompt engineering approach, wherein the model was primed with a structured instruction set designed to align its internal weights with the specific linguistic patterns of academic abstracts. This priming mechanism involved assigning the model a specific role to narrow its focus to formal scientific terminology, supplemented by exemplar-based guidance. By providing diverse examples for each category, such as "Finite Element Method" for methodologies, the prompt calibrated the model's understanding of the required granularity without requiring a full fine-tuning process. To ensure high data integrity, strict syntactic constraints were applied to preserve the original surface forms of entities as they appeared within the proposal text. Additionally, the model was instructed to decompose entities containing stopwords into smaller text sub-chunks, ensuring that the resulting data remained clean and accessible for downstream analyses.

Beyond granular entity extraction, the pipeline performed a coarse-grained topic classification requiring the model to assign each proposal to one of 26 high-level fields derived from OpenAlex Topics. This dual-task approach facilitated a hierarchical analysis of the dataset, allowing for both broad disciplinary mapping and specific conceptual analyses. The comprehensive prompt utilised for this extraction and classification process, including the full list of OpenAlex categories, is documented in Appendix A.

*3.3. Topic Classification*

To support our macro-level analysis of grant proposals, we performed topic classification to map each submission to its most semantically aligned research area within a standardised taxonomy. We utilised an off-the-shelf product developed by the Innovation Growth Lab in collaboration with DSIT, namely the *DSIT-Taxonomies* (Innovation Growth Lab, 2025). This tool categorises UKRI-funded research by linking project descriptions to recognised classification systems, including *OpenAlex Topics*, employing a blend of semantic similarity analysis and natural language inference to suggest labels and associated confidence scores.
The OpenAlex Topics framework is structured as a four-level hierarchy:

**domain → field → subfield → topic**

Under the DSIT-Taxonomies approach, a proposal is assigned a path that begins at the domain level but may terminate at any stage of the hierarchy. While this provides relevance scores linking proposals to specific taxonomy nodes, the raw assignments often result in heterogeneous granularity. Relying solely on these inconsistent assignments would compromise fine-grained resolution and limit the depth of downstream statistical analyses. To address this, we implemented a refinement procedure to extend all hierarchical assignments down to the most granular topic level.
The DSIT-Taxonomies tool operates via a two-stage pipeline. It first extracts research entities using standard libraries, such as DBpedia Spotlight (Mendes, 2011), RAKE (Rose, 2010), YAKE (Campos, 2020), and KeyBERT (Grootendorst, 2020). Then, it classifies proposals by mapping these entities to the most relevant OpenAlex Topics. Our approach refines this process, by replacing the initial extraction stage with the domain-specific, Mistral-based procedure outlined in Section 3.2, while retaining the original classification module.
Building on this, we further developed its classification algorithm by enforcing a mandatory downward-traversal constraint. This ensures that every proposal is mapped to a leaf topic node, as illustrated in Figure 2, thereby mitigating the granularity imbalance inherent in the raw algorithmic output and providing a uniform resolution for downstream statistical analysis.

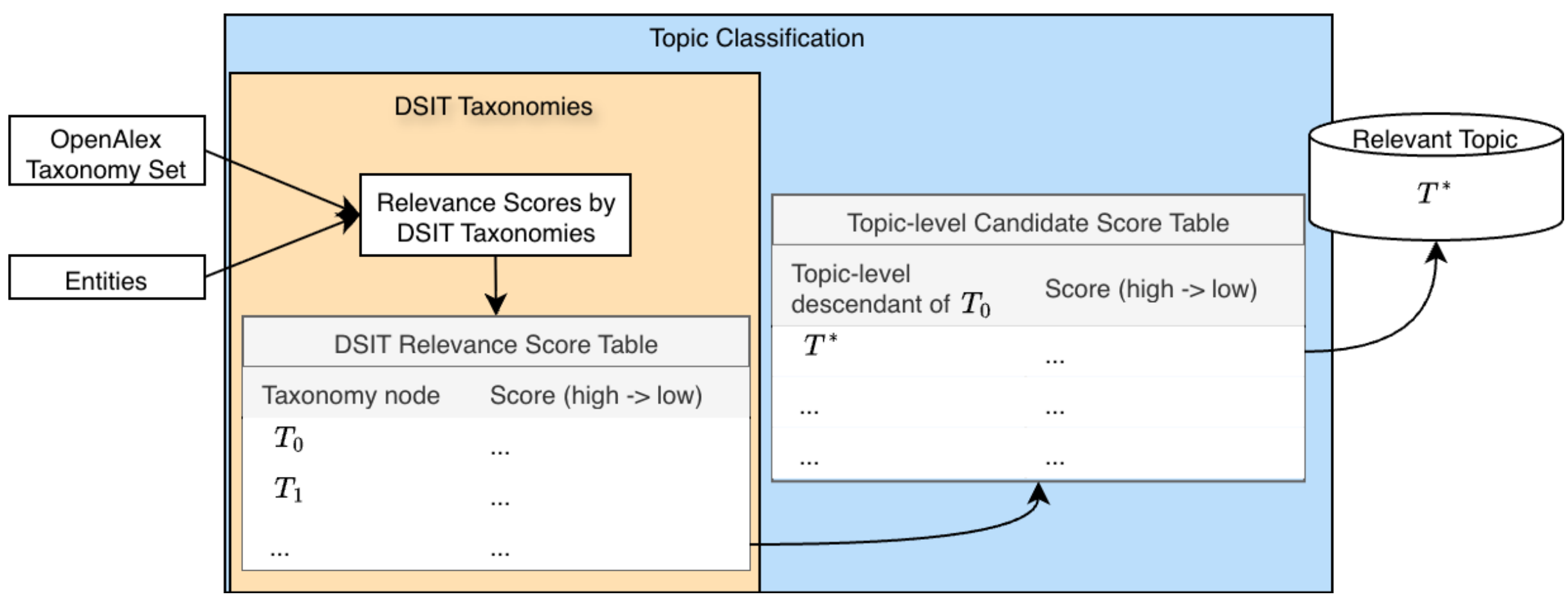

**Figure 2: Workflow of the Topic Classification.**

In practice, let $T_0$ denote the DSIT-Taxonomies-assigned taxonomy node for a proposal *p*. This node may correspond to a domain, field, subfield, or topic within OpenAlex Topics. Instead of accepting $T_0$ as final, our algorithm restricted the DSIT-Taxonomies relevance score table to all topic-level descendants of $T_0$, which was denoted as $Desc(T_0)$. This produced a topic-level candidate score table, containing only terminal nodes beneath $T_0$ and their associated relevance scores. The proposal was then classified as

$$T^* = arg_{t \in Desc(T_0)} max\ Score(t, p)$$

Effectively, this procedure transformed coarse DSIT-Taxonomies assignments into fully resolved topic-level labels while remaining consistent with the original scoring structure.

### *3.4. Experimental SetUp*

Experiments were conducted using a semi-automated Python pipeline executed in a Jupyter Notebook, ensuring a consistent workflow. The database was constructed by randomly sampling 42 projects from the GtR database. Mistral and DSIT-Taxonomies were run locally, and GPT-4o was accessed via its API.

## 4. Results

### *4.1. Evaluation Dataset*

For the evaluation, 42 proposal applications were randomly selected, ensuring a broad distribution across the various source councils and different time periods, to capture a wide range of academic disciplines and evolving research themes. This selection process also accounted for variations in text length, with the sample's combined title and abstract character counts ranging from 404 to 4,049 (averaging 2,698). As no pre-existing gold-standard benchmark exists for this specific extraction task, the evaluation relied on intensive manual assessment. Two authors (XR and AS) independently reviewed the outputs for these 42 proposals, assessing both the relevance of extracted entities and the correctness of topic assignments. Any disagreements were resolved through discussion and consensus. This rigorous human-in-the-loop process necessitated a dataset size that remains manageable for experts while being statistically significant for meaningful analysis. By ensuring this sample is highly diverse, we have constructed a saturated dataset, ensuring our findings remain valid and generalisable across the wider corpus of grant proposals.

### *4.2. Baselines*

To ensure a rigorous assessment of our pipeline, we established distinct baselines for both the entity extraction and topic classification stages. For the Research Entity Extraction evaluation, we benchmarked the performance of our Mistral-based approach (Sec. 3.2) against two comparative standards: the GPT-4o LLM and the DSIT-Taxonomies algorithmic approach. To maintain experimental consistency between the LLM-based methods, we utilised an identical prompt template for both GPT-4o and Mistral. Conversely, the DSIT-Taxonomies approach was employed as a standalone, off-the-shelf tool, representing the existing non-generative methodology.

For the Topic Classification evaluation, we examined the accuracy of the customised DSIT-Taxonomies classification logic (Sec. 3.3) when supplied with two different input sets: entities extracted via our Mistral-based pipeline and those retrieved through the standard DSIT-Taxonomies baseline. This dual-input approach allows us to determine whether the generative extraction method provides a more effective foundation for downstream classification than traditional algorithmic extraction.

### *4.3. Research Entity Extraction Evaluation*

We compared the performance across three approaches (Mistral, GPT-4o, and DSIT-Taxonomies) using a metric-based evaluation to assess statistical performance, and through a comparative content analysis to systematically characterise the nature and quality of the extracted entities. The latter involved a structured review of the annotated entities to classify error types and assess contextual relevance.

#### *4.3.1. Metric-Based Evaluation*

This evaluation focused on the distribution and overlap of the entity sets for each proposal. We first examined the volume of entities generated by each method. Table 1 details the average, maximum, and minimum number of entities identified within our sample. The data reveals that the two LLM-based approaches (Mistral and GPT-4o) yield a comparable average number of entities, whereas the DSIT-Taxonomies algorithm produces significantly more, averaging roughly 10 additional entities per proposal.

**Table 1: Descriptive statistics of entities extracted per approach.**

| *Approach* | *Average Entities* | *Max Entities* | *Min Entities* |
|---|---|---|---|
| ***Mistral*** | 15.94 | 31 | 0 |
| ***GPT-4o*** | 14.3 | 46 | 6 |
| ***DSIT-Taxonomies*** | 25.22 | 43 | 10 |

Notably, the minimum count of zero observed for Mistral was attributed to a model hallucination, or when information is shown to be present in the outcomes, but not in the input text or supported by the data, during the extraction process. This specific instance highlighted the necessity of developing more resilient algorithms and future iterations of this work should include automated retry logic to mitigate such anomalies and ensure robust data capture.

To assess the consistency between methods, we measured the extent to which the entity sets overlapped across the proposals. We performed pairwise comparisons by calculating the Jaccard Similarity for each proposal, defined as:

$$Jaccard(A, B) = \frac{|A \cap B|}{|A \cup B|}$$

The Jaccard index ranges from 0 (no overlap) to 1 (identical sets). While this matching can be performed syntactically, we opted for semantic matching. Syntactic comparison is often too rigid. For example, it would treat “stem cells” and “stem cell” as a mismatch despite them representing the same research concept (Salatino et al., 2019).

To facilitate semantic matching, all entities were converted into 384-dimensional dense vector embeddings using the “all-MiniLM-L6-v2”[1] Sentence-BERT model. These embeddings map semantically similar concepts to proximal locations in a multidimensional space, ensuring that minor variations (such as pluralisation or hyphenation) do not result in a false mismatch. We utilised cosine similarity to determine the proximity of entities, employing a threshold of 0.6 to identify semantically equivalent terms. Table 2 presents the average Jaccard similarity scores derived from these pairwise analyses.

**Table 2: Pairwise Jaccard similarity (Semantic) across approaches.**

| | **DSIT-Taxonomies** | **GPT-4o** | **Mistral** |
|---|---|---|---|
| **DSIT-Taxonomies** | - | 0.179 | 0.192 |
| **GPT-4o** | 0.179 | - | 0.374 |
| **Mistral** | 0.192 | 0.374 | - |

[1] Sentence Transformers https://huggingface.co/sentence-transformers/all-MiniLM-L6-v2

The results indicate that Mistral and GPT-4o produce the most similar entity sets, with a Jaccard index score of 0.374. Conversely, the DSIT-Taxonomies algorithm shows the lowest alignment with the LLM-based approaches. This analysis suggests that while Mistral and GPT-4o generate comparable outputs, DSIT-Taxonomies operates on a distinct extraction logic that results in a significantly different entity profile, hence the low Jaccard index of 0.192.

### *4.3.2. Comparative Content Analysis*

To assess the quality of the extracted entities, we developed a bespoke interface within a Jupyter Notebook. Figure 3 shows this interface displaying each research proposal with entity annotations from all three methods overlaid as purple bounding boxes. This visual framework allows assessors to directly compare the contextual relevance and precision of each approach.

Annotated Text

Entities extracted using **DSIT**

--------

Centre for Stem Cell Biology and Medicine. Every year, thousands of patients die from organ failure because there are not enough matching donors for the transplants that they need. Also, thousands of others suffer from the effects of diabetes, even though they take insulin. Existing therapies for many diseases of the brain, such as Parkinson?s and multiple sclerosis, fall far short of a cure. Stem cells could possibly provide cures for these diseases. Embryonic stem (ES) cells are derived from embryos at the earliest stages of development, when the embryo is a microscopic, hollow ball of cells. They provide a promising means for generating many or perhaps all of the 200 specialised cell types of the body. Stem cells have been detected in other body tissues, such as the bone marrow and nervous system, and these also represent potential sources of therapeutic materials. The proposed research is designed to provide the foundations in basic knowledge of stem cell biology to build the clinical treatments that are likely to emerge from a fundamental understanding of this field.

--------

Annotated Text

Entities extracted using **GPT**

--------

Centre for Stem Cell Biology and Medicine. Every year, thousands of patients die from organ failure because there are not enough matching donors for the transplants that they need. Also, thousands of others suffer from the effects of diabetes, even though they take insulin. Existing therapies for many diseases of the brain, such as Parkinson?s and multiple sclerosis, fall far short of a cure. Stem cells could possibly provide cures for these diseases. Embryonic stem (ES) cells are derived from embryos at the earliest stages of development, when the embryo is a microscopic, hollow ball of cells. They provide a promising means for generating many or perhaps all of the 200 specialised cell types of the body. Stem cells have been detected in other body tissues, such as the bone marrow and nervous system, and these also represent potential sources of therapeutic materials. The proposed research is designed to provide the foundations in basic knowledge of stem cell biology to build the clinical treatments that are likely to emerge from a fundamental understanding of this field.

--------

Annotated Text

Entities extracted using **MISTRAL**

--------

Centre for Stem Cell Biology and Medicine. Every year, thousands of patients die from organ failure because there are not enough matching donors for the transplants that they need. Also, thousands of others suffer from the effects of diabetes, even though they take insulin. Existing therapies for many diseases of the brain, such as Parkinson?s and multiple sclerosis, fall far short of a cure. Stem cells could possibly provide cures for these diseases. Embryonic stem (ES) cells are derived from embryos at the earliest stages of development, when the embryo is a microscopic, hollow ball of cells. They provide a promising means for generating many or perhaps all of the 200 specialised cell types of the body. Stem cells have been detected in other body tissues, such as the bone marrow and nervous system, and these also represent potential sources of therapeutic materials. The proposed research is designed to provide the foundations in basic knowledge of stem cell biology to build the clinical treatments that are likely to emerge from a fundamental understanding of this field.

--------

**Figure 3: Interface developed via Jupyter Notebook, displaying comparative annotations for DSIT-Taxonomies (left), GPT-4o (middle), and Mistral (right). The green box displays the approach.**

A primary observation from our analysis is the fragmented nature of the DSIT-Taxonomies approach. As shown in the left panel of Figure 2, DSIT-Taxonomies frequently identifies generic terminology but fails to capture complex, specialised concepts, particularly those formed by English compound nouns. For example, the term “signal amplifier” was incorrectly extracted as two distinct entities, failing to recognise that “signal” acts as a qualifier for “amplifier”. Moreover, the algorithm often extracts non-substantive elements, including verbs (e.g., “provide”), quantifiers (e.g., “thousands”), and adverbs (e.g., “however”). This tendency introduces noise, rendering the DSIT-Taxonomies approach less suitable for identifying high-quality research entities, and even more so if these entities change over time.

Consistent with the previous analysis in Sec. 4.3.1, GPT-4o and Mistral yield comparable results but differences between them are subtle. We observed occasional instances where Mistral extracted fewer entities than GPT-4o and vice versa.

We also examined the prevalence of hallucinations, a known challenge for generative models where entities are suggested that do not exist in the source text. Our observations indicate that Mistral is more conservative in this regard, producing fewer non-existent entities than GPT-4o, maintaining a significantly lower mean error rate per proposal (1.56 compared to 2.84).

Furthermore, Mistral recorded 16 proposals with zero hallucinations, whereas GPT-4o reached this benchmark in only 7 cases. Since these hallucinations do not appear in the original text, they are easily identifiable and can be effectively mitigated during post-processing via a routine syntactic verification that ensures all extracted entities are grounded in the source document.

### *4.4. Topic Classification Evaluation*

To evaluate the topic classification stage, we utilised the customised DSIT-Taxonomies classification module to map entities to defined research topics. We compared the performance of this module when using two different input sources: i) entities extracted via our Mistral-based approach and ii) those extracted by the original DSIT-Taxonomies algorithm.

Also in this process, to facilitate human assessment, we developed a Jupyter Notebook interface that displayed the original proposal alongside extracted entities and resulting topics side-by-side, as shown in Figure 4. This allowed annotators to assess which input set yielded the most contextually appropriate research topic for each application.

| Annotations | Council | Topic using Mistral Entities | Topic via DSIT Entities |
|---|---|---|---|
| Annotated Text<br>Entities extracted using MISTRAL<br>--------<br>Divining Roots: uncovering how SUMO-mediated responses control developmental plasticity. Food security represents a major global issue. Significant improvements in crop yields are urgently required to meet the increase in world population by 2050. The ability of a crop to efficiently absorb water and nutrients from soil is dependent on its root system responding to the availability of these resources. For example, roots preferentially form branches when in contact with water employing a mechanism called hydropatterning. Understanding the regulation of root branching is of vital agronomic importance.<br>This research project will investigate how hydropatterning works to position new root branches in response to water availability. Our research project attempts to 'fill in the gaps' between roots sensing water availability and then branching. To help our studies, we have already identified plant signals and genes such as auxin and ARF7 that are important for this process. Several promising processes will also be characterised including one that modifies ARF7 that switches on root branching. The knowledge gained from this study will provide new information about the key genes and processes controlling root branching in response to water availability, helping scientists design novel approaches to manipulate root architecture to enhance resource capture and yield in crops.<br>-------- | BBSRC | Plant and Biological Electrophysiology Studies (level 4)<br>Plant Science (level 3)<br>Agricultural and Biological Sciences (level 2)<br>Life Sciences (level 1) | Tree Root and Stability Studies (level 4)<br>Mechanical Engineering (level 3)<br>Engineering (level 2)<br>Physical Sciences (level 1) |

**Figure 4: Interface developed via Jupyter Notebook, displaying the annotated proposal, and the OpenAlex topics extracted using the research entities extracted using Mistral and the DSIT-Taxonomies algorithm.**

Table 3 shows that with 90.5% (38 correct classifications and 4 errors), the Mistral-based approach significantly outperformed the full DSIT-Taxonomies pipeline attaining 71.4% correct classifications (30 correct classifications and 12 errors).

**Table 3: Topic classification accuracy across 42 grant proposals.**

| | *Correct* | *Errors* | *% Correct* |
|---|---|---|---|
| ***Mistral*** | 38 | 4 | 0.905 |
| ***DSIT-Taxonomies*** | 30 | 12 | 0.714 |

## 5. Discussion and conclusions

The results of our evaluation suggest that while Mistral and GPT-4o demonstrate high performance in entity extraction, they are most effectively deployed at scale (i.e., for large datasets which limit human evaluation). By processing thousands of proposals, the aggregate dataset achieves a comprehensive level of thematic coverage that naturally offsets minor individual inaccuracies. This allows a robust foundation for research mapping, where the collective reliability of the research entities provides an actionable overview of the research landscape.
From an operational perspective, Mistral offers a distinct advantage due to its lightweight architecture compared to GPT-4o (OpenAI, 2026). This efficiency is particularly relevant for institutional deployments, as Mistral can be integrated within secure, internal platforms. For government organisations, like UKRI, handling confidential grant applications, the ability to execute high-performance extraction within a protected environment ensures data sovereignty while maintaining the analytical depth required for complex research classifications.
It is also important to acknowledge that the entity extraction evaluation measures the inter-model alignment, rather than performance against a human-verified standard. Whereas the evaluation of our topic classification provides a definitive measure of accuracy, establishing the objective correctness of the identified topics.
Our next step involves applying this approach to the full UKRI dataset, comprising 350K of both funded and unfunded proposals. By leveraging OpenAlex Topics as thematic containers, we will perform advanced trend detection to analyse the evolution of the UK research landscape over time, specifically benchmarking domestic progress against global competitors. In parallel, we will utilise extracted research entities and established prediction techniques (Salatino et al. 2018) to conduct fine-grained analyses and identify emerging trends. To ensure a comprehensive view, we will extend our classification beyond titles and abstracts to include secondary outputs such as “Impact Summaries” and “Key Findings”.

**Open science practices**

The analyses presented in this manuscript fully adhere to the open science practices. The dataset and the code is publicly available under a Creative Commons Attribution 4.0 International License (CC BY 4.0), and it can be downloaded from: https://doi.org/10.5281/zenodo.18937183.

**Acknowledgments**

We would like to thank the UK Metascience Unit and all the participants of the Data Sandpit for Metascience (https://www.ukri.org/opportunity/data-sandpit-for-metascience/) who helped shape the idea behind this project.

**Author contributions**

The authors contributed to the work as follows, according to the CRediT system:
**Conceptualization**: Xingran Ruan, Angelo Salatino; **Methodology**: Xingran Ruan, Angelo Salatino; **Software**: Xingran Ruan, Angelo Salatino; **Validation**: Xingran Ruan, Angelo Salatino; **Writing - Original Draft**: Xingran Ruan, Angelo Salatino, Gemma Derrick; **Writing - Review & Editing**: Xingran Ruan, Angelo Salatino, Rosa Filgueira, Kara Morav, Alexandru Marcoci, Gemma Derrick, Sarah Callaghan; **Funding Acquisition**: Angelo Salatino, Rosa Filgueira, Alexandru Marcoci, Gemma Derrick, Sarah Callaghan

**Funding information**

This research is part of the “Tracking Stars and Unicorns” project funded by the UK Metascience Unit, established in 2024 as a joint unit led by the Department for Science, Innovation and Technology (DSIT) and UK Research and Innovation (UKRI). More info: https://www.gov.uk/government/publications/a-year-in-metascience-2025

**Competing interests**
No competing interests.

## Appendix A

Here we report the prompt we used to extract entities from the grant proposals:

```
In this prompt, you will receive a title and abstract of a research proposal.
Your task is twofold. First parse the content in the <research_proposal> tag,
identify and extract some relevant entities:

- research topics (E.g., "social impact of technology", "urban heat island
effect", "patient adherence to medication", "long-term memory formation")
- discipline (E.g., "Computational Linguistics", "developmental psychology",
"Materials Science", "public health sector")
- methodologies (E.g., "quantitative analysis", "Finite Element Method", "Gas
Chromatography", "content analysis", "statistical modelling")
- approaches (E.g., "case study", "experimental design", "mixed-methods
approach", "systematic review", "longitudinal analysis")
- tools (E.g., "mass spectrometer", "telescope arrays", "Scanning Electron
Microscope", "High Performance Computing cluster")
- dataset (E.g., "demographic census data", "patient electronic health records",
"financial market time series", "interview transcripts")
- software (E.g., "R statistical package", "AutoCAD", "MATLAB", "secure
blockchain platform", "NVivo for qualitative coding")
- artefacts (E.g., "validated survey instrument", "working prototype",
"simulation model", "policy recommendation report", "code repository",
"technical specification")
- events/occurrences (E.g., "2008 financial crisis", "Mount Vesuvius eruption",
"COVID-19 pandemic", "the Meiji Restoration", "September 11")


Constraints for Extraction:
1. Return the entity in the same form it appears in the '<research_proposal>'.
This is important for preservation.
2. If the entity contains stopwords, break it down, and return the text
subchunks.


Your second task is to identify the most appropriate field of research for this
proposal according to the following list of 26 fields:

Earth and Planetary Sciences; Physics and Astronomy; Business, Management and
Accounting; Agricultural and Biological Sciences; Environmental Science;
Economics, Econometrics and Finance; Social Sciences; Medicine; Arts and
Humanities; Biochemistry, Genetics and Molecular Biology; Chemistry;
Engineering; Computer Science; Energy; Immunology and Microbiology;
Neuroscience; Decision Sciences; Mathematics; Materials Science; Psychology;
Dentistry; Health Professions; Chemical Engineering; Nursing; Pharmacology,
Toxicology and Pharmaceutics; Veterinary

<research_proposal>
{title and abstract of the proposal}
</research_proposal>
```